# Graph Computing for Financial Crime and Fraud Detection: Trends, Challenges and Outlook

Eren Kurshan

*Columbia University*
*New York, NY, 10027, U.S.A.,*
*ek2925@columbia.edu*

Hongda Shen

*University of Alabama*
*Huntsville, AL 35899, U.S.A.*
*ahs0017@alumni.uah.edu*

The rise of digital payments has caused consequential changes in the financial crime landscape. As a result, traditional fraud detection approaches such as rule-based systems have largely become ineffective. AI and machine learning solutions using graph computing principles have gained significant interest in recent years. Graph-based techniques provide unique solution opportunities for financial crime detection. However, implementing such solutions at industrial-scale in real-time financial transaction processing systems has brought numerous application challenges to light. In this paper, we discuss the implementation difficulties current and next-generation graph solutions face. Furthermore, financial crime and digital payments trends indicate emerging challenges in the continued effectiveness of the detection techniques. We analyze the threat landscape and argue that it provides key insights for developing graph-based solutions.

## 1. Introduction

Digital payments have experienced an unparalleled growth in the past decade [1],[2],[3]. In 2019 alone, 743 million transactions (valued at 187 billion USD) were processed through the Zelle digital payment network alone. This translates to a 57% year-to-year growth in the total transaction amounts and 72% increase in the transaction volumes [4]. Between 2013-2018, mobile payments grew by over 120% (in compounded annual growth rate) in China [1]. Similarly in India digital payments volume rose by 61% over the past 5 years [5].

Globally, mobile banking and digital payments have provided billions of people the opportunity to access financial services. Furthermore, they have delivered practical benefits to individual consumers, businesses and financial service providers, such as time savings, speed, ease of use, lower transaction costs and the ability to scale [6].

On the other hand, criminal schemes have rapidly evolved to benefit from the new fast-moving digital payments landscape. Traditionally, fraud and financial crime detection relied on a large number of rules and static thresholds to flag suspicious transactions (e.g. amounts larger than $10,000). In the recent past, such manual and rule based techniques have become ineffective as fraudsters rapidly figure out the static rules and bypass them.

Global financial crime volume was estimated to be around 1.4-3.5 trillion USD per year according to the latest industry reports [7]. Other sources estimate money laundering to be around 2-5% of the global GDP (up to 1.87 trillion EU), a large percentage of which is not detected [8]. 3.2 million fraud records were filed through the U.S. Federal Trade Commission (FTC) System in 2019 alone; indicating a 53% increase from 2018 [9].

The cost to fight and recover from fraud has also increased by over 30% since 2016 [10]. In addition to the increases in the fraud cases, fraud schemes also changed considerably. In the past few years, with the general adoption of the EMV credit and debit cards, in-person card fraud has declined sharply [11]. Yet, a substantial surge has been reported in the attempted online fraud [12]. As of 2020, online credit card fraud cases rose due to the global pandemic and the corresponding increases in the digital transaction volumes [13].

Financial crime has negative effects on individuals and financial institutions as well as systemic effects such as negative consequences on a countries welfare through macroeconomic performance. In the past few years, AI and machine learning solutions have gained considerable interest in the financial services industry, including the compliance and risk management functions [14]. Graph algorithms and databases have long been considered important tools in fraud detection[15]. Numerous studies have demonstrated the effective use of anomaly detection, network flow and sub-graph based analysis [16, 17]. Lately, graph neural networks have gained interest [18], [19], [20], [21]. Prior to the financial services applications, graph neural networks have been implemented in a wide range of industries. Their broader potential to improve generalizations and relational reasoning has been highlighted by a number of studies [22, ?], [23]. In financial crime and fraud detection they provide enhanced performance and flexibility [23].

However, implementing graph-based solutions in real-life transaction processing and crime detection systems brings new challenges to light. In this paper, we take a practical look into the use of graph computing in financial crime detection applications. We highlight the difficulties development organizations face in building and deploying graph-based solutions in financial transaction processing systems. Furthermore, we analyze the emerging financial crime trends and discuss their implications on the detection techniques. Both financial crime detection and graph computing are very broad and rapidly evolving fields. The purpose of this paper is not to provide a comprehensive overview of the corresponding fields (as it should be understood that both will change quickly), but to highlight the general painpoints and overarching practical implementation considerations in order to develop more effective solutions. This paper is organized as follows: Section 2 is a high-level overview of the characteristics of financial crime types as well as recent trends; Section 2.5 presents highlights from the state-of-the-art research in graph-based solutions; Section 3 reviews the application considerations; and finally Section 4 provides conclusion discussions and outlook.

## 2. Background

This section provides an overview of some of the common financial crime and fraud types. It discusses the main characteristics and emerging trends to identify the requirements for the next-generation graph-based detection techniques. Recent financial crime trends underscore the constantly changing nature of the field and emphasize the need for innovative solutions.

### 2.1. *Financial Crime Types and Trends*

Financial crime is defined as any non-violent crime that generally results in a financial loss [24]. According to [25] it can be categorized into four main types: corruption, fraud, theft and manipulation. This paper focuses on the last three categories and the corresponding detection techniques. Financial crime schemes have been going through major changes lately [26],[27]. According to the recent industry surveys [28], almost all fraud types have seen serious increases. External fraud has been rising, both in terms of volume and total transaction amounts by 61% and 59% respectively.

*1. Payment Fraud*

Payments fraud covers criminal activity in numerous payment channels including credit and debit card transactions, ATM, person-to-person (P2P) transactions, wire, automated clearing house transactions, online payments, automated bill payments, checks and deposits. In the past few years, pervasive increases have been seen in fraud across all payment channels [27].

According to J.P.Morgan and the Association for Financial Professionals, payments fraud is currently reported at over 80% of organizations and continues an upward trend [29]. Card fraud (including credit and debit card fraud) is one of the largest segments in payments fraud, with losses that amounted to 1.8 billion EUR in 2016 [30] according to the European Central Bank. Chip cards and PIN usage has caused shifts in the fraud tactics from card-present transactions (CP) to card-not-present (CNP) transactions [31]. Despite the introduction of 3D-Secure, CNP fraud is growing and amount to over 70% of the total card fraud losses. All online payment channels (from bill payments to P2P and online check processing) experienced significant growth in recent years.

In addition to the rapid move to the online channels and the changes in the fraud tactics, frequent cross-overs among payment channels and fraud types have been reported. As an example, the World Bank reports indicate that criminals increasingly utilize mobile payment channels for money laundering [32]. Since the economies of scale appeal to financial crime organizations, they continue to pursue large merchant hacks to gain access to card related information for tens or hundreds of millions of accounts [31],[33], which fuels the online fraud.

*2. Identity Theft*

Identity theft refers to the process of stealing an individual's personal information to commit fraud. ID theft schemes use an ever-changing list of tactics ranging from ATM skimming devices to phishing, smishing, dumpster diving, and compromised wireless networks. Following the identity theft itself, perpetrators typically use the compromised information across multiple channels. Identity theft and new account frauds cause more financial damage compared to the other payment fraud types due to the time it takes to detect them. As reported by [34], close to 58% of the ID theft cases are discovered after 4 months. Further, approximately 25% of the cases are discovered after 2 years. Such long discovery periods are advantageous to the financial crime organizations and motivate further investments in ID theft schemes.

Since 2000, there has been a steady growth in the number of identity theft reports and complaints [34]. Lately, ID theft has become one of the top fraud types in the Federal Trade Commission criminal filings [9]. Identity theft is typically associated with a collection of other crime types [34]. Credit card fraud was the top fraud type for downstream fraud in 2019, with over 200K cases filed. New credit account fraud grew by 88% during the same period [9]. Likewise, check fraud and financial scams are frequently associated with ID theft [34].

*3. Financial Scams*

Financial scams refer to a wide range of fraud and criminal activities that use deception as a means to steal from targeted individuals. These crimes use continuously evolving tactics such as phone scams, technical support scams, elderly scams, charity and lottery scams, ticket scams etc. [35]. Each category then includes a long list of tactics, such as customized elderly scams involving insurance, IRS impersonation scams, mortgage, and grandparent scams etc. [35].

Due to the increase in the elderly population, financial scams that target the elderly have become more prominent in recent years. The Federal Bureau of Investigation reports that 80% of all scams in the U.S. target at-risk populations between the ages of 55–85 [36]. Financial scams usually tie to identity theft and account takeover, as well as their downstream fraud types in multiple channels. Financial scams have become one of the top concerns in the fraud landscape [9]. Analyzing interconnectivity and shared network patterns are key in detecting financial scams and the downstream fraud cases.

*4. Account Takeover*

Account takeover fraud occurs when perpetrators gain access to a victim's bank account illegally. During ATO, criminals typically also change the account login credentials and contact information, so that the victim is not able to access the account. They eventually drain the funds through one or more payment channels (such as wire transfers). ATO has strong ties to cybersecurity as perpetrators regularly use mass data breaches, mobile SIM hijacking, compromised devices and networks to fuel their attacks. Improved authentication technologies, end-point security and other cybersecurity practices play a key role [37]. Akin to identity theft, ATO provides a gateway to numerous downstream fraud types.

ATO has experienced a 78% increase in 2019 alone as reported in [38] (other studies report bigger increases). Furthermore, recent news reports indicate that the incidents of ATO skyrocketed between Q1-Q3 during the global pandemic in 2020. Similar alarming rates were reported between 2019-2020 year-to-year.

Recently, the crime topologies as well as the channels through which crime is being committed have been changing rapidly [39]. Financial crimes, fraud and cybersecurity threats are rapidly converging [40]. ATO provides a prime example of this convergence and the challenges of detecting criminal activity. Cybercrime tactics like spear-phishing, back-doors, infected devices, ATM hacks, balance alterations have become increasingly prominent in the financial services attacks. This further threatens the traditional financial services defense mechanisms that are gradually becoming ineffective in coping with such changes and convergence. On the other hand, since financial crime organizations seek the efficiency in repetitive attacks, this provides significant opportunities in fraud detection through graph-based techniques and the integration of cyber-defenses with financial crime detection.

*5. Synthetic ID and Account Fraud*

Synthetic account fraud is based on fabricated identities made to look like real customers with favorable credit scores and characteristics. Synthetic ID fraud cases regularly use social security numbers and credit privacy numbers blended with real and synthetic information from one or more individuals.

In recent years, there has been a significant growth in the synthetic ID and account fraud. Latest reports highlight that the fraud type has grown by around 35% year-to-year [41]. One of the reasons for this trend is the fact that synthetic ID/account fraud is its non-transactional nature. Hence, it takes longer to discover and usually does not have the traditional crime reporting paths other fraud types rely on (e.g. reporting fraudulent credit card transaction). Hence, the resulting financial gains for the criminal organizations are proportionally higher due to the longer time-to-discovery.

*6. Money Laundering*

Money laundering is defined as the transfer of illegally obtained money or funds generated by criminal activity (such as drug trafficking, terrorist financing etc.) to conceal the true source [24]. Money laundering can occur with or without the knowledge of the financial firms or the counter-parties involved in the financial transactions.

Money laundering typically consists of 3 main stages each of which require customized techniques for detection: (i) *Placement:* Proceeds from the illegal activities are placed into the financial system. The funds are frequently structured into smaller amounts, placed into multiple accounts, in a number of financial institutions to prevent detection. Criminals typically make small cash deposits, purchase monetary instruments/bank drafts in order to remain under the radar for reporting based on the anti-money laundering laws and regulations [42],[43], [44]. (ii) *Layering:* During the layering stage, multiple transfers occur between shell companies and individuals to conceal the origin of the illicit funds. Anti-money laundering *(AML)*, know-your-client *(KYC)* and countering financing of terrorism *(CTF)* have been highly critical functions to prevent such transfers in financial institutions. However, anti-money laundering efforts have been facing serious difficulties over the past few years. (iii) *Integration:* During the integration stage, the funds are integrated into the financial system as if they were legal. This stage may involve investments (such as real estate or securities), purchases of luxury items etc.

Globally, an estimated 715 billion to 1.87 trillion EUR is laundered according to the industry reports [8]. Over the years, the number and the variety of transactions used in money laundering have become increasingly complex. They often involve a large number of accounts, transaction types, numerous financial institutions from many jurisdictions, and growing use of non-bank financial institutions (such as check cashing services, insurance etc.). This complexity makes detection efforts significantly more challenging.

*7. Other Fraud Types*

It is important to note that fraud and financial crimes span a large spectrum of criminal activities. In this paper, we highlight a selection based on their prominence and the emerging trends. Mortgage fraud and loan scams have been quite prominent in the overall fraud landscape. They generally involve customer account or personal identifying information (PII) compromises, often tied to identity theft. Hence, both the customers and the financial institutions suffer losses. Despite its importance in the overall fraud landscape, mortgage fraud has experienced decreases over the past few years.

Internal fraud involves financial services employees engaging in criminal activities including but not limited to financial sanction breaches, insider trading, bid rigging, price fixing, market allocation and bribery. As highlighted earlier with its examples, external fraud can be categorized as first-party fraud (involving the bank's own client) or third-party fraud (involving external contacts and criminal organizations). Internal fraud solutions also rely on entity-based interconnectivity analysis used for first and third-party external fraud for detection. Both financial losses and regulatory compliance play as important motivating factors in detecting internal fraud.

## 2.2. *Shared Characteristics*

As noted earlier, financial crime tactics have been displaying high-degrees of adaptiveness to the emerging digital landscape, significant convergence with cybersecurity-threats and and increased propensity for efficiency mechanisms. As an example, the rise of identity theft and ATO and decline in in-person (CP) credit card fraud highlights this efficiency drive. Crime tactics rapidly adapt to payment processing system vulnerabilities as well as the prevention measures.

Fraud tactics show remarkable levels of customization to the individual channels and the multi-channel modes. Channels display notable differences in fraud characteristics, such as transaction type, amounts, processing times, devices, authentication requirements etc. For example, ATM fraud is considerably different than online bill payment fraud (in terms of frequency, amounts, transaction and processing time-ranges, parties involved, access compromises, devices involved etc.). These unique characteristics play an important role in the effectiveness of the algorithmic solutions. In many cases the techniques that perform well in one channel face performance challenges in others due to such differences (e.g. how dynamic the environment is, scale, attributes etc.). Furthermore, the ability to balance the channel-specific patterns with the need to integrate the techniques for multi-channel fraud remains a challenge.

According to Interpol, most financial crimes are of cross-border, trans-national nature, through the use of digital platforms and internet [45]. In addition to the increased ability to conceal their identities over diverse geographical footprints, financial crime organizations rely on complex transaction patterns and diversified geographical operations. Transnational fraud also provides the criminal organizations the ability to transfer illegal funds to locations with reduced scrutiny. While online fraud remains to be a significant challenge, graph-computing provides significant advantages in achieving such interconnectivity-based analysis at a global scale.

### 2.3. *Roles of Financial Institutions*

Financial firms play a key role in detection and prevention of financial crime. The Treasury Department has estimated that 99.9% of the criminal money that is presented for deposit in the United States gets into secure accounts [46].

A financial institution can be involved in financial crime as victim, as perpetrator, or as an instrumentality. (i) *Victim:* Financial firms are frequently faced with fraud losses (such as misrepresentation of information in synthetic identity credit and loan applications). Third-party fraud is one of the largest portion of the financial crimes in terms of volumes and monetary value. (ii) *Perpetrator:* As noted in Section II.A. financial firms may be directly involved in the financial crimes through internal fraud, self dealing, collusion, misappropriation schemes etc. (iii) *Instrument:* Financial institutions may also serve as intermediaries and instruments in transferring funds and performing transactions, knowingly or unknowingly. Money laundering frequently involves banks as instruments. Though the liabilities may vary depending on the financial crime type, financial firms have been faced with significant fraud losses and have made significant investments in the recent past to upgrade their detection and prevention systems.

**2.4.** *Financial Crime Laws and Regulations*

Over the years, numerous laws and regulations have been enacted for financial crimes. The Bank Secrecy Act (1970) established the first set of record-keeping and reporting requirements by individuals, banks and other financial firms to prevent money laundering. Its main goal was to assess the source, volume and flux of currency and monetary instruments in/out of the U.S.A or in financial institutions. This requires the financial institutions to (i) Report any cash transaction above $10K; (ii) Identify the parties; (iii) Generate a paper record trail.

Bank Secrecy Act has been one of the key tools to combat money laundering. However, until 2012 BSA required Suspicious Activity Reports (SARs) that are filed to the Office of the Comptroller of the Currency (OCC) were not digitized and had manual processing components, which posed as an obstacle, and was enhanced through electronic filing systems that improved the speed and efficiency of the processing. Later, Money Laundering Control Act (1986) established money laundering as a federal crime. Anti-Drug Abuse Act of 1988 extended the scope from banks and financial institutions to cover real estate, automotive dealers in the requirements for reporting the large currency transactions (over $3K).

The Patriot Act (2001) and Intelligence Reform Terrorism Prevention Act (2004) extended the Bank Secrecy Act by including terrorist financing as a crime, prohibited financial institutions to engage with foreign shell banks, enhanced due-diligence and know your client procedures and finally enhanced the cross-institution information sharing among the government agencies and the financial institutions. Finally, Intelligence Reform Terrorism Prevention Act of 2004 expanded the Bank Secrecy Act with regulations requiring cross-border electronic transmittal of funds. Most of these laws provided valuable data and data acquisition mechanisms that are essential for the detection solutions to run on.

In addition to the financial crime laws, financial firms and transactions are subject to numerous regulations that cover many aspects from fairness and discrimination to data quality, explainability and interpretability to cybersecurity requirements [47], [48], [49], [50], [51]. As an example, a credit card transaction, loan and mortgage application and other financial transactions may be subject to Consumer Credit Protection Act, Fair Credit Reporting Act, Equal Credit Opportunity Act, Fair and Accurate Credit Transactions Act, as well as regulatory directives such as Regulation B. GDPR [52] and CCPA [53] require explainability, transparency and interpretability conditions as well as data related requirements. Even though laws and regulations do not appear critical from a solution development perspective, such regulations are imperative in designing the solution approach. They highlight the fundamental requirements (such as interpretability, transparency etc.) data sources, quality requirements and operational boundaries.

**2.5. *Graph-based Detection Techniques***

2.5.1. *Traditional AI and Machine Learning*

Machine learning has been used in fraud detection since the 90s [54], [55],[56],[57]. Over the years, graph and network analysis techniques have been established as important tools in both research and industrial practice [15]. Graphs inherently exhibit advantages in representing the underlying financial transaction data. The nodes and edges often represent companies, individuals, accounts, transfer of funds, locations, devices, and other financial or non-financial data. Depending on the application, a diverse range of graph types (directed, undirected, cyclical, acyclical, static, dynamic, attributed and colored) have effectively been utilized.

2.5.2. *Data Mining*

Early data mining techniques for fraud detection relied on neural networks, regression, support vector machines, Bayesian networks that operated on tabular representations [56], [58], [59], [60]. Later, graph-based representations have been explored [61, 16] such as community of interest selection [62], dynamic graphs [63], and signature-based systems [64].

2.5.3. *Graph Anomaly Detection*

Anomaly detection has been deployed in financial services systems to capture the differentiating characteristics of fraud in the immense quantities of financial transaction data [65], [66]. It provides insights on the data patterns by focusing on the distinct characteristics in terms of connectivity, entity characteristics, flow, traffic patterns (both locally and globally), sub-graph characteristics and events in the graph representations [67, 68], [69],[66].

2.5.4. *Supervised and Sub-Graph Analysis*

Sub-graph analysis and mining analyzes the local graph patterns through supervised as well as semi- or unsupervised learning [70], [71]. In addition to the connectivity, traditional network flow techniques (e.g. min-cut/max-flow) help identify the sub-graphs of interest. As an example, sub-graphs analysis may identify small-scale fraud rings generating abnormal patterns over a large number of accounts. Nevertheless, as fraud patterns are highly dynamic and adversarial, the use of a fraud detection technique causes changes in the fraud tactics to prevent detection. For instance, the use of strongly connected components by graph-based detection algorithms has motivated perpetrators to hide their activities by artificially creating networks to camouflage their activities. The use of metric-based analysis to detect such cases has been explored by some researchers [70].

2.5.5. *Flow and Path Analysis*

In money laundering, the process of *layering* directs the flow of the illegal funds through multiple parties to prevent detection. One of the limitations of the dense sub-graph based techniques is that they mostly focus on single-step transfers. Hence, they face difficulties and require adjustments to detect multi-stage cases frequently used in money laundering. Recently, multi-partite and multi-step solutions [72], flow analysis and k-step neighborhood based techniques [73] have been proposed to address this. Network flow solutions, traditionally used for intrusion detection, are also of great interest for financial crime detection use cases.

2.5.6. *Graph-based Machine Learning*

In recent years, graph-based ML and graph neural networks have gained significant interest [20, 74]. Graph neural networks provide advantages over traditional neural networks in providing better generalizations and improved relational reasoning. Their applications in financial crime detection have yielded promising results [21], [75]. Despite being in the early stages of exploration in fraud detection, GNNs provide flexible representations both in terms of attributes and graph structures. They enable a level of structural configurability along with the ability to compose architectures containing multiple blocks. Graph neural networks can operate on a variety of graph types and numerous studies demonstrated GNNs in financial crime detection. [76] used graph convolutional networks (GCN) to detect money laundering. [77] explored dynamic considerations in graph networks, which is essential in fraud applications. [78] proposed solutions to expose camouflaged fraudsters through GNNs. Later studies aimed to improve the structural limitations of the graph networks through attention mechanisms for higher efficiency [79].

**2.6.** *Graph Use Cases*

Graph techniques are used in different aspects of fraud detection:

2.6.1. *Transaction and Event Scoring*

Graph techniques are commonly used in transaction fraud risk scoring and detecting suspicious events. Risk scoring and detection are often conducted in real-time while the transaction is being processed. The results are compared to predetermined risk thresholds and may trigger alerts for financial crime related events accordingly.

2.6.2. *Entity Scoring*

Though it is not as well known, graph computing techniques are also used for Black and Gray-List generation as well as tracking fraud rings. In these applications, graph techniques are utilized to score the risk scores of entities based on their historical profiles and transaction events (including monetary and non-monetary transactions such as account access/logins etc). The entities that are scored may be accounts, clients as well as entities such as devices such as computers and mobile phones, local area networks, automated teller machines etc. The entities deemed high risk may be placed on black-lists to prevent further transactions that involve them or may be added to tracking lists to continue monitoring their behavior.

### 2.6.3. *Sub-graph Scoring*

Beyond the individual nodes, groups of nodes and sub-graphs may be analyzed and tracked to uncover the financial crime organizations through transaction characteristics as well as through the profiling of the sub-graphs. Similar to the entity scoring, sub-graph scoring can often occur offline independent of the real-time transaction processing restrictions that affect the transaction scoring.

### 2.6.4. *Operational Use Cases and Alert Processing*

Though not as commonly known, graph computing techniques are used in operational use cases in fraud processing systems. One common example is the alert processing. Fraud and financial crime detection models often generate a large number of fraud alerts during their normal operation. These alerts are processed through various channels, including fraud analysts and automated alert processing systems (e.g. text messages and mobile alerts). Depending on the use case, alert processing may occur either during or after the transaction processing.

For high-throughput channels such as card payment transactions alert processing occurs after an approve/decline decision has been made by the transaction processing system. This is in contrast to the other channels, such as wire transactions, automated clearing house (ACH) and online/bill payments, for which alert processing may range from end-of-business day to the deadline of the transaction itself. This changes the temporal characteristics of the processing task significantly. However, for most channels, due to the large number of alerts generated, manual alert processing systems may face bandwidth limitations [80]. Prioritizing the accounts based on the risk assessment through graph analysis provides a solution towards lower financial losses and improved response times.

### 2.6.5. *Crime Investigations*

During alert and suspicious alert processing human fraud and financial crime analysis work together on investigations. Investigations are enhanced through graph visualizations as well as graph-based event analysis. Similarly, graph computing techniques provide opportunities to enhance alerts by improving the interpretability. To summarize, graph computing provides a large number of solution opportunities during the end-to-end financial crime detection, investigation and prevention process.

## 3. Application Considerations

This section discusses some of the common practical implementation considerations and pain points from a model development perspective.

### 3.1. *Temporal Considerations*

A large portion of the financial crime and fraud detection solutions is implemented within the transaction processing systems. These complex systems process large amounts of transaction data in real-time. Transaction processing systems typically have millisecond-range response time SLAs, including the time to process the transaction itself (accessing the account information, checking the availability of the funds), fraud scoring, payment network processing, communication protocols etc. Graph computing solutions face serious response time pressures due to the combination of real-time processing constraints and the use of large, interconnected graphs. In the past decade, traditional solutions relied on memory efficient graph representations, graph compression techniques [81] [76] and distributed graph computing to deal with the large graphs in financial use cases [82],[83].

### 3.2. *Pattern Extraction through Attention*

Attention techniques have gained increased popularity in the past years, due to their natural effectiveness fraud detection [84]. Fraud patterns often subtle by design, to prevent detection by authorities. Many fraud rings practice well-planned *account priming* periods to conceal the transaction patterns with similar patterns that make them appear more normal. A common example of this in ATO cases, is to send small amounts to the perpetrator's account that may not trigger alerts yet will make the larger transfers that follow seem normal by building history. Graph attention has been utilized to detect such patterns in account takeover [85].

The placement of funds during money laundering involves hard to detect transaction patterns performed by multiple entities. The goal is to transfer funds between the carefully concealed source and sink nodes through various transaction channels [42]. Similarly, during the structuring phase, funds are divided into smaller amounts to prevent reporting. Attention techniques become vital to detect such cases and identify the important patterns even though they are buried in large transaction volumes. Similarly, graph attention networks eliminate the requirement of knowing the graph structure in advance and enables focusing on the most relevant parts of the input [79]. Fraud applications frequently borrow techniques cybersecurity space (such as intrusion detection) in response to the agility pressures. Optimizing the system for data updates, selecting specialized temporal features,and cost sensitive learning have been borrowed and explored from intrusion detection solutions [86, ?],[87]. Other studies adapt the graph convolutional network (GCN) models along the temporal dimension to deal with the dynamic complexities.

**3.3. *Dynamic Graphs and Data Updates***

Globally, medium to large banks serve tens of millions of clients on a daily basis over multiple processing channels. Payments channels are interconnected through entities (such as clients and accounts), hence the transaction throughput translates to tens of millions of updates to the underlying graph representations. High-frequency transaction types (e.g. P2P payments and credit card payments) require more frequent updates than slower channels (e.g. bill payments). Though they are in early stages, recent techniques focus on the dynamic nature of the transaction systems through dynamic graph neural networks [88].

During the deployment of the graph systems, both nodes (representing the entities such as account holders) and edges (representing the relationships such as transfer of funds) may be updated. Detecting anomalies in constantly and rapidly changing graphs at industrial scale is a significant research problem. Recently, techniques have been proposed to detect anomalies in real-time or near real-time in order to determine if an incoming edge is anomalous as soon as it is received [89]. Anomaly detection for such graphs with transient characteristics can be done at the node-level [90], at the edge-level [90], as well as at the sub-graph level [91]. Evolving graph neural networks [77], streaming GNNs [92], temporal networks [93] have also been proposed to perform deep learning on dynamic graphs. Though in earlier stages, these solutions target a range of link durations and update frequencies.

**3.4. *Scale and Complexity Issues***

Financial transaction processing systems are typically complex and involve numerous payment transaction types and models to process fraud risk and business strategies.

3.4.1. *Number, Size and Variety of Graphs*

In payment systems, the features of interest for fraud detection models may reside in different reference graphs of different types and disparate characteristics (both internal to the financial institution and from external sources). This complexity causes difficulties in developing effective graph solutions. Thanks to their structural characteristics, graph neural networks have parallelism potential, which, in theory, makes it feasible to perform batch computations from independent graphs. Recently, techniques have been proposed to provide the ability to operate on diverse graph types [79]. Likewise, cross-channel fraud requires computations on multiple graphs with different characteristics simultaneously under the pressures of real-time response service-level agreements (SLAs). Though limited, some solution approaches have been proposed to integrate multiple graphs in unified structures [94] and perform label propagation [95],[96], yet no clear solution has emerged to address the practical challenges.

Large graphs, such as those used in the financial crime and fraud detection appli-cations, pose unique challenges due to their sheer size. In transaction graphs tens of millions of customers need to be represented in payment systems, with similar size and complexity in the interconnectivity. Extracting insights on the large graph itself or the sub-graphs within, remain as challenges. Representing the differences in the characteristics of different sub-graphs is essential in identifying financial crime. As an example, a collection of nodes that appear as a clique in the graph may represent an ill-concealed money laundering fraud ring. In the past few years, techniques to categorize and describe the differences in different graph regions have been proposed [97]. The value proposition of such techniques is higher in cases where the interpretation of different sub-graphs is needed for investigations and alert resolution purposes on large graphs.

### 3.4.2. *Entity and Feature-level Complexity*

Financial transaction data usually includes the parties involved in the transactions, flow of funds, historical connectivity between parties, cross channel events, risk profiles and many other features. In card payment transactions, hundreds of features are passed from the payment networks during the transaction processing and merged with a large number of features internally. Depending on the use case, a large number of complex features may be used in graph-based techniques and calculated in real-time to reach approve/deny decisions in a timely manner.

### 3.4.3. *Algorithmic Complexity and Response Time*

While not emphasized in most studies, algorithmic complexity is also of interest for large scale and time-sensitive implementations such as fraud detection. Graph neural networks have been shown to be effective on much larger networks than what they have been trained on, which is an advantage in large scale implementations [98]. Still, the combination of size and algorithmic complexity drives serious implementation challenges.

### 3.4.4. *Increasing Fraud Complexity*

As noted in the earlier sections, fraud schemes have been growing increasingly complex over the years. The complexity is evident in almost all dimensions, including the number of channels involved, the activity and events patterns, temporal characteristics, entities involved etc. As an example, account takeover fraud frequently involves a large number of subtle non-monetary events that occur in numerous channels, including phone calls to customer service call centers, online non-monetary transactions (such as account setting changes), account access and login events (such as one-time-password (OTP) requests), monetary transactions (such as card payments, wires, person-to-person transactions) etc.

During the detection process, the events that occur over long periods of time (sometimes months) are evaluated with respect to the large number of entities and their risk profiles. The goal is to identify the risk and point of account compromise (such as malware on the device used for login). Later, the transactions, events and entities (such as accounts, devices) need to be marked for processing. As discussed in Section II, even traditionally single-channel fraud schemes have gradually shifted to complex, multi-channel events. This creates complexity both in the detection algorithms, the number and complexity of the graphs and graph techniques that are involved. It also requires seamless transitioning between graphs, platforms, transaction processing systems, channels etc. as well as integration to detect anomalous patterns across multiple channels. Individual stages of the financial crimes may involve different transaction and payment systems, require disparate detection strategies and techniques, while the siloed solutions need to be integrated to compose a coherent view and analyze the suspicious activity across different platforms and systems.

### 3.5. *Data Quality, Complexity and Noise*

In addition to the naturally occurring noise in the data, in fraud use cases the criminals often attempt to inject deceptive data into the financial transaction processing systems to conceal their activities. In synthetic account fraud, account takeover and other crimes, such data may become a determining factor in the decision making processes. While there are mitigation strategies for such attacks, robustness and performance of the underlying algorithmic solutions under data quality pressures are rarely considered in research studies. Financial crime detection solutions use diverse data types and data sources like unstructured data from images, video, audio recordings, text, emails, structured data from financial transactions, account information and historical records. However, continuous data integration and update processes, compounded with the data quality and complexity issues create a demanding environment and require targeted solutions.

### 3.6. *Labeled Data Availability and Infrastructure Support*

While most transaction-based detection systems (such as card payments) have well-established and automated infrastructure support for labeled data generation, other use cases vary in terms of the effort required to generate such data. Depending on the use cases, the labeled data limitations may also be caused by the low fraud rates, the lack of reporting or the timely integration of the data. Even though anomaly detection solutions exhibit natural advantages in such cases with limited labeled data, improving the performance of unsupervised and semi-supervised solutions has been challenging.

### 3.7. *Algorithmic Standards, Benchmarks and Tools*

As highlighted earlier, graphs algorithms have been gaining popularity in many application domains including financial services. Standardization, benchmarking and tool efforts are useful for both academic applications and industrial deployment systems [99]. The underlying diversity of the primitive building blocks of graph algorithms may be a challenge as different groups create different overlapping variants instead of developing common low level building blocks. Recently, there have been efforts to develop standard set of primitive building blocks for graph techniques [100]. Similarly, benchmarking frameworks for graph neural networks to add new models for arbitrary datasets have been proposed recently [101]. Likewise, from a computational perspective benchmark frameworks have been developed to improve the efficiency of graph algorithms [102]. Though financial applications are yet to be included in such benchmarks the continuous improvement efforts will likely include financial service customizations in the near future.

### 3.8. *Adversarial Nature and Robustness Issues*

Financial crimes are classic examples of dynamic adversarial applications. Their adaptiveness can be in the form of:

(i) *Bypassing Prevention Measures:* As discussed in Section 1, rule-based fraud detection approaches have largely become ineffectively as perpetrators quickly estimate the static rules and develop techniques to bypass them. Supervised learning techniques also face serious challenges, as fraudsters often adopt novel techniques to avoid machine learning based detection. As an example, when fraudulent payment transactions from a financial crime organization are declined by the machine learning based detection systems, novel tactics are often utilized immediately.

(ii) *Channel Usage and Fraud Scheme Shifts:* As noted earlier, financial crimes topologies have largely shifted from simpler, single channel cases to, complex multi-channel cases. Likewise, criminals have been utilizing novel channels that are not traditionally associated with the corresponding fraud types. As an example, industrial reports claim that emerging channels like bitcoin transactions have been gaining popularity for money laundering activities [103], [32].

(iii) *Rapidly Changing Tactics over Time:* Mobile person-to-person (P2P) payment fraud is known to exhibit abrupt tactical shifts, which creates a challenging environment for supervised learning based solutions. Anomaly detection, and adaptive algorithms have been explored and deployed for these use cases. In the long term, adaptive and adversarial techniques provide the most promising solution path forward as they reflect the application environment most accurately.

(iv) *Robustness:* Adversarial tactics also play an important role in the overall robustness of the detection solutions. Often, the criminal organizations have the ability to react to the preventative measures rapidly and at-scale. Their increased technological investments and automation capabilities are a growing concern [104]. In practical systems, these capabilities can have serious implications on the robustness of the machine learning based detection models.

### 3.9. *Increased Automation and Machine Learning Attacks*

Machine learning attacks are emerging concerns as they target highly-critical applications. Numerous of studies illustrated that ML models may face robustness issues during possible poisoning, evasion, and inference attacks [105], [106], [107], [108]. Neural networks are known to be sensitive to even small perturbations in the data [109] and hence are vulnerable to adversarial attacks. Building robust models has become a major implementation goal in industrial use cases [110], [111]. Building strategies against poisoning attacks and other machine learning attack types have gained significant interest in the past few years [112], [113], [114]. At this point, the research on robustness and adversarial challenges for graph-based machine learning systems is in its nascent stages [115].

The fact that fraud rings increasingly rely on automation and advanced technologies [104], highlights the fundamental limits of success for the current and next-generation fraud prevention techniques. Agility, automation and scale become critically important in the deployment systems. Graph computing solutions need to quickly adapt to the growing challenges of robustness and ML attacks.

### 3.10. *Governance and Regulatory Compliance*

While most financial crime laws and regulations involve documentation and reporting procedures, model governance processes in the financial firms present a challenging environment for machine learning and graph-based techniques. As showcased in [116], financial institutions often require conservative modeling approaches towards model performance guarantees for compliance purposes. As a result, many promising techniques experience issues during the model governance reviews (due to their inherent risks and the lack of performance guarantees for compliance).

Beyond the approval challenges, the model governance and risk management procedures take significantly longer for novel techniques. This extends the time-to-deploy and put the model development timelines and detection performance at risk. As an example, reinforcement learning provides key advantages in adversarial environments [117] thanks to its origins in gaming. Graph reinforcement learning studies have shown promising results in financial crime detection use cases [118], [119]. Yet, their implementation in the financial services industry has been slower. Similarly, autonomous learning approaches and online algorithms are of great interest for fraud detection [120]. Nevertheless, the underlying autonomy and the lack of performance guarantees become issues for the internal regulatory compliance reviews. Current model governance processes require revisions and adjustments to speed up the adoption of more autonomous machine learning approaches for fraud detection.

### 3.11. *Interpretability and Explainability*

Recent regulations like CCPA [53] and GDPR [52] impose interpretability and explainability mandates on most AI/ML models in financial services. Yet, expanding the capabilities of the current explainability solutions has been an industry-wide challenge [121], [122], [123], [124]. Explainability and interpretability become even more challenging in large dynamic graphs that are used in financial crime detection applications. Beyond regulatory requirements explainability techniques are needed for audit purposes and inquiries for declined transactions as well as for crime investigations. Lately, some research studies have explored these challenges in dynamic graphs [125]. Developing optimized explainability solutions for emerging graph models and satisfy the regulatory requirements are highly critical [126],[127], [128].

### 3.12. *Visualizations and Crime Investigations*

Visualizations are used for multiple purposes in crime detection systems, including but not limited to tracking emerging fraud trends, performing criminal investigations, investigating fraud alerts and collaborating with other fraud analysts. Although they have been effectively used in other fields, conventional visualization and human-in-the-loop tools face serious obstacles in fraud detection. Graph computing provides intrinsic advantages in representing and visualizing the data. Yet, scaling the solutions to the industrial scales and dealing with the speed and complexity challenges are serious issues [129],[130]. In the recent past, large graph visualization solutions were benchmarked around tens of thousands of node range. Later super-graph visualizations were proposed for hundreds of thousand nodes [131]. However, financial transactions visualizations require graph visualizations for graphs in hundreds of millions, which still poses serious challenges [132]. Similarly, the interactivity requirements, graph complexity, number of graphs involved and interpretability issues plague the graph-computing solutions and the corresponding visualizations.

### 3.13. *Integrating Emerging Techniques and Data Sources*

Emerging approaches that involve behavioral profiling and non-monetary events provide promising results. These solutions rely on large amounts of data from various hybrid data sources, such as click-streams, logins, page views, mobile device profiles etc. Integrating such solutions and the hybrid data sources with traditional graph-based representations and data types in complex transaction processing systems is an emerging challenge.

### 4. Conclusions and Outlook

In this paper, we overview the common application challenges in graph-based fraud and financial crime detection systems. Financial crime and fraud schemes have exhibited rapid changes in the past few years. Driven by the growth in the digital payments landscape, crime topologies and characteristics (from the transaction types to the channels and tools) have transformed, which left traditional detection solutions ineffective. Even machine learning techniques have been facing issues in dealing with the increasingly subtle fraud patterns. Graph techniques provide intrinsic advantages in detecting fraud in large volumes of digital transaction data [133]. Both traditional graph computing solutions and emerging techniques are of great interest from an industrial deployment perspective.

However, implementing and deploying graph computing techniques in real-life detection systems pose unique difficulties, due to the size, speed, complexity and adversarial characteristics of the financial crime detection applications. In addition, the underlying complexity of the digital transaction processing systems, including the large-scale implementation requirements, real-time processing, siloed nature of the channels, frequent updates, complex data/graphs, makes the deployments and reaching the detection performance targets difficult.

While graph neural networks and emerging adaptive solutions provide important tools to shape the future of fraud and financial crime detection, the highly-adaptive nature of fraud will likely continue posing challenges. As highlighted throughout the paper, the convergence of financial crime, fraud and cybersecurity threats is a significant concern for the financial services industry due to the siloed and disparate nature of the corresponding systems. Graph-based solutions and systems are essential to both customize and integrate the numerous detection and prevention systems. The paper argues that focusing on the application-specific requirements provides opportunities to improve the current and emerging graph-based solutions.